\documentclass[prd,aps,preprintnumbers,fleqn,showpacs,nofootinbib,superscriptaddress]{revtex4}
\usepackage{amssymb,amsmath,graphicx,bm}

\usepackage[usenames,dvipsnames]{color}
\usepackage[normalem]{ulem} % \sout{old text} for strikeout
\renewcommand\sout{\bgroup \color[rgb]{0.55,0.00,0.99} \ULdepth=-.5ex \ULset}

\newcommand{\nn}{\nonumber}
\newcommand{\beqn}{\begin{eqnarray}}
\newcommand{\eeqn}{\end{eqnarray}}

\newcommand{\st}{{\scriptscriptstyle T}}
\newcommand{\sperp}{{\scriptscriptstyle \perp}}
\def\nn{\nonumber}
\begin{document}

%\preprint{NIKHEF 2015}

\title{Energy evolution of T-odd gluon TMDs at small $x$}

\author{Dani\"el Boer}
 \affiliation{ {Van Swinderen Institute for
Particle Physics and Gravity, University of Groningen, Nijenborgh 4,
9747 AG Groningen, The Netherlands}}

\author{Yoshikazu Hagiwara}
 \affiliation{ Key Laboratory of
Particle Physics and Particle Irradiation (MOE),Institute of
Frontier and Interdisciplinary Science, Shandong University,
(QingDao), Shandong 266237, China }

\author{Jian Zhou}
 \affiliation{ Key Laboratory of
Particle Physics and Particle Irradiation (MOE),Institute of
Frontier and Interdisciplinary Science, Shandong University,
(QingDao), Shandong 266237, China }

\author{Ya-jin Zhou}
\affiliation{ Key Laboratory of
Particle Physics and Particle Irradiation (MOE),Institute of
Frontier and Interdisciplinary Science, Shandong University,
(QingDao), Shandong 266237, China }

\begin{abstract}
  We study the energy or TMD evolution of the three leading twist dipole type T-odd gluon TMDs inside a transversely polarized nucleon, all of which at small $x$ dynamically originate from the spin dependent odderon. Their energy dependence presents a unique opportunity to study the polarization dependent TMD evolution in the small-$x$ region, where the distributions are identical up to a normalization constant at tree level. We further propose to study their evolution via azimuthal asymmetries in virtual photon-jet production in polarized proton-proton collisions at RHIC. We present model predictions for the asymmetries as functions of the large jet or photon transverse momentum and $Q^2$ which set the hard scales in this process.
  \end{abstract}
  
  % \pacs{...}
  \date{\today}
  
  \maketitle
  
  \section{Introduction}
  Three dimensional tomography of gluonic matter inside a nucleon/nucleus has become a topic of much interest in recent years, in large part due to the increasing prospects of collider experiments at small-$x$ values. Transverse momentum dependent (TMD) gluon distributions play a central role in describing the 3D structure of nucleon/nucleus in momentum space. In particular, a very rich phenomenology of polarization dependent TMDs in the small-$x$ limit has been uncovered in a series of work~\cite{Metz:2011wb,Dominguez:2011br,Zhou:2013gsa,Szymanowski:2016mbq,Schafer:2013opa,Dumitru:2015gaa,Boer:2015pni,Hatta:2016dxp,Zhou:2016rnt,Iancu:2017fzn,Cougoulic:2019aja,Kovchegov:2015pbl,Kovchegov:2020hgb,Cougoulic:2020tbc,Boussarie:2019icw,Mantysaari:2019csc,Dumitru:2021tvw,Kovchegov:2021iyc,Bhattacharya:2022vvo,Hagiwara:2021xkf}. The theoretical and experimental studies of polarized gluon TMDs at small $x$ are of great importance not only because they encode information on the nucleon/nucleus internal structure, but also because they allow to address many interesting aspects of QCD dynamics, for instance, the (non-)universality of TMDs, the gluon polarization of the Color Glass Condensate (CGC), and the properties of the elusive odderon that has finally been discovered recently \cite{Abazov:2020rus}. 
  
It was recognized some time ago that at small $x$ two different definitions of the unpolarized gluon distributions play a role \cite{Kharzeev:2003wz}. Depending on the process that is being considered one probes different TMDs that are distinguished by their different gauge link structures: the dipole or Weizs\"acker-Williams distributions, a mixture of both or even more complicated options~\cite{Dominguez:2010xd,Dominguez:2011wm,Bomhof:2006dp}. In this work, we focus on the dipole type gluon TMD which has a closed loop gauge link in the fundamental representation and can be related to gluonic correlators that appear in the description of saturation phenomena at small $x$.  The relations between small $x$ correlators and gluon TMDs allow to study gluon TMDs by employing the powerful theoretical tools established  in saturation physics. Some interesting developments along this line include the computation of the linearly polarized gluon TMD in the McLerran-Venugopalan (MV) model~\cite{Metz:2011wb}, and deriving a TMD resummation formalism based on the CGC effective theory~\cite{Mueller:2013wwa,Zhou:2016tfe,Xiao:2017yya,Zhou:2018lfq}. A more recent finding~\cite{Boer:2015pni,Kovchegov:2021iyc} reveals that the three leading twist dipole type T-odd gluon TMDs (including the gluon Sivers function) inside a transversely polarized nucleon share a common dynamical origin and become identical at small $x$. All three distributions are related to the spin dependent odderon which not only plays a role in generating transverse single spin asymmetries in high energy scattering~\cite{Zhou:2013gsa,Dong:2018wsp,Yao:2018vcg,Santiago:2021shh,Kovchegov:2021iyc}, but also contributes to the polarization averaged scattering amplitudes~\cite{Boussarie:2019vmk,Hagiwara:2020mqb}. 
  
  The relation among the three T-odd TMDs only holds at tree level, but is spoiled by higher order radiative corrections. This is to be expected because the resummation is formulated in impact parameter space and different Bessel functions and weights appear due to the different polarization tensor structures associated with these T-odd gluon TMDs. Since these distributions cannot be calculated from first principles, it will have to be checked experimentally in how far the equality holds or not, but it can be calculated how the functions change w.r.t.\ each other under energy evolution. We will consider this using the 
 diquark model expressions from~\cite{Szymanowski:2016mbq} as starting distributions and for one particular process. 
  At present there are only few known processes that probe the dipole gluon TMDs, one of which is virtual photon-jet production in proton-proton or proton-nucleus collisions, which in the forward or backward region is dominated by the $q+g \rightarrow \gamma^*+{\rm jet}$ process. In \cite{Boer:2017xpy} we studied azimuthal asymmetries in this process for unpolarized collisions. In the present paper we will study polarized proton-proton collisions which allow to study the three dipole type T-odd gluon TMDs inside a transversely polarized nucleon and their energy evolution through azimuthal spin asymmetries. Assuming that the three gluon TMDs are the same at the initial scale, our numerical estimations indicate that the equivalence of these distributions is significantly violated fairly quickly at higher scales and that the differences in azimuthal asymmetries might be visible in experiments at RHIC using the STAR forward detector.

  We thus propose to access the T-odd gluon TMDs by measuring the azimuthal angle dependent cross section of virtual photon plus jet production in polarized proton-proton collisions. All three T-odd gluon TMDs induce different azimuthal modulations and can therefore be fully analyzed in this single process. We restrict to the backward region of the transversely polarized proton, such that a hybrid approach~\cite{Gelis:2002ki} can apply, in which one uses a collinear PDF on the unpolarized proton side while multiple gluon scatterings on the polarized proton side are treated in the CGC formalism. Such a hybrid approach remains a good approximation as long as the typical transverse momenta of the incoming gluon inside the CGC are much larger than that carried by the incoming quark. In contrast, for virtual photon-jet production at mid rapidity and/or at larger $x$ values, 
  the hybrid approach would not be applicable and effects from color entanglement in the full TMD description would need to be taken into account, thereby reducing or possibly even removing the predictive power of the QCD calculation~\cite{Collins:2007nk,Rogers:2010dm} (see also relevant discussions~\cite{Schafer:2014xpa,Zhou:2017mpw,Buffing:2018ggv}). For the phenomenological applications of this hybrid formalism, we refer readers to Refs.~\cite{Akcakaya:2012si,Schafer:2014xpa,Kotko:2015ura,Boer:2017xpy,Marquet:2017xwy,Dumitru:2018kuw,vanHameren:2020rqt,Altinoluk:2019fui,Fujii:2020bkl,Altinoluk:2021ygv}.

TMD factorization at  small-$x$  can only be established in the so-called correlation limit~\cite{Dominguez:2010xd,Dominguez:2011wm}. To reliably extract small-$x$ T-odd gluon TMDs, we will therefore focus on the correlation limit in virtual photon-jet production in polarized proton-proton collisions, which means that the outgoing virtual photon and jet are approximately back-to-back in transverse momentum. 
  
  The paper is structured as follows. In Sec.II.\ we present detailed numerical estimations of the evolved T-odd gluon TMDs using the diquark model result for the spin dependent odderon as the input at the initial scale. We derive the polarization and azimuthal angle dependent cross section for virtual photon-jet production, and make predictions for RHIC energy in Sec.III. The paper is summarized in Sec.IV.

\section{Evolved T-odd gluon  TMDs}
We start with introducing the matrix element definition of gluon TMDs.  The information on confined gluon transverse motion inside a transversely polarized target is formally encoded in the following correlator matrix element,
\begin{eqnarray}
\Gamma^{\mu \nu [U,U']}(x,\bm{k}_{\perp}^2) = \frac{1}{xP^+}\int \frac{dy^-
d^2y_T}{(2\pi)^3} e^{ik \cdot y} \langle P, \bm{S}_\perp | 2 {\rm Tr} \left [
F_{T}^{+\mu}(0) U F_{T}^{+\nu}(y) U' \right ] |P, \bm{S}_\perp \rangle
\big|_{y^+=0} , \label{gmat}
\end{eqnarray}
where $U$ and $U'$ are process dependent gauge links in the
fundamental representation. $\bm{S}_\perp$ is the nucleon transverse spin vector. One can define six leading power gluon TMDs by
parameterizing the tensor structure of the above correlator~\cite{Mulders:2000sh, Boer:2016xqr},
\begin{eqnarray} \nn
   \Gamma^{ij}(x,\bm{k}_\perp^2) &=&   - \,g_\st^{ij} \,f_1^g(x,\bm{k}_\perp^2) + \frac{k_\perp^{ij}}{M^2} \,h_1^{\perp g}(x,\bm{k}_\perp^2)  - \,\frac{g_\st^{ij} \epsilon_\st^{S_\perp k_\perp}}{M} \,f_{1T}^{\perp g}(x,\bm{k}_\perp^2) + \frac{i \epsilon_\st^{ij} \bm{k}_\perp \cdot \bm{S}_\perp}{M} \,g_{1T}^g (x,\bm{k}_\perp^2) \\ &&
    - \,\frac{\epsilon_\st^{k_\perp\{i} S_\perp^{j\}} + \epsilon_\st^{S_\perp\{i} k_\perp^{j\}}}{4M} \,h_1^g(x,\bm{k}_\perp^2) - \frac{{\epsilon_\st^{\{i}}_\alpha k_\perp^{j\}\alpha S_\perp}}{2M^3} \,h_{1T}^{\perp g}(x,\bm{k}_\perp^2)  ,
 \label{gg_para}
\end{eqnarray}
where the six gluon TMDs are functions of $x$ and $\bm{k}_\perp^2$,
$\epsilon_T^{\mu \nu}=\epsilon^{\rho \sigma \mu \nu} p_\rho
n_\sigma$ with $\epsilon^{-+12} = 1$, $g_T^{\mu\nu}=g^{\mu\nu}-p^{ \{\mu} n^{\mu \} } /p \cdot n$, 
and $k_\perp^{i_1 \ldots i_n}$ are completely symmetric and traceless tensors which up to rank $n=3$ are given 
by $k_\perp^{ij} \equiv \;k_\perp^i k_\perp^j + \frac{1}{2} \bm{k}_\perp^2 g_\st^{ij} $ and $ k_\perp^{ijk} \equiv \;k_\perp^i k_\perp^j k_\perp^k + \frac{1}{4} \bm{k}_\perp^2 \left( g_\st^{ij} k_\perp^k + g_\st^{ik} k_\perp^j + g_\st^{jk} k_\perp^i \right) $.

The first two gluon TMDs, $f_1^g$ and $h_1^{\perp g}$, are the
unpolarized and linearly polarized gluon distributions, respectively.   Among the four transverse spin dependent gluon TMDs, the three T-odd gluon TMDs, $f_{1T}^{\perp g}$, $h_{1T}^{\perp g}$ and $h_{1}^{g}$, are relevant for the single spin asymmetry studies. Depending on the process considered, there are two main types of gauge link structure appearing in the gluon distributions: the Weizs\"{a}cker-Williams (WW) distribution and the dipole type distribution. The former has a staple like gauge link, while the latter contains a closed loop gauge link in either the adjoint representation or the fundamental representation. The unpolarized gluon distribution $f_1^g$ and linearly polarized gluon TMD $h_1^{\perp g}$ have the same $1/x$ enhancement in the small-$x$ limit for both the WW and the dipole cases~\cite{Metz:2011wb,Dominguez:2011br}. It has been found that the  WW type T-odd gluon TMDs and the dipole type T-odd  distributions with a gauge link in the adjoint representation  are suppressed in the small-$x$ limit~\cite{Schafer:2013opa,Boer:2016fqd}.  In contrast, all of the three dipole type T-odd gluon TMDs with a gauge link in the fundamental representation rise rapidly with decreasing $x$ (although not as rapid as the unpolarized gluon distribution ~\cite{Hatta:2005as,Kovchegov:2003dm,Bartels:1999yt}, thanks to a $1/x$ enhancement obtained in higher order calculations). 

Interestingly, at tree level the three dipole type T-odd gluon TMDs can be related to the spin dependent odderon~\cite{Boer:2015pni,Kovchegov:2021iyc,Boer:2016xqr},
 \begin{eqnarray}
  x f_{1T}^{\perp g}
  = x h_1^g
  = - \frac{\bm{k}_\perp^2}{2M^2} x h_{1T}^{\perp g} 
  =\frac{\bm{k}_\perp^2 N_c}{4 \pi^2 \alpha_s  }O_{1T}^\perp
 \label{todd} \,.
\end{eqnarray}
 The derivation of the above relation is in close analogy to that of the following  relation~\cite{Metz:2011wb,Dominguez:2011br},
 \begin{eqnarray}
x f_1^g = \frac{\bm{k}_\perp^2}{2M^2} x h_1^{\perp g},
\label{linearly}
\end{eqnarray}
where the unpolarized and the linearly polarized gluon TMDs are the dipole type distributions. These relations remain true under small-$x$ evolution. In the QED case, the same relation between the unpolarized photon TMD and the linearly polarized photon TMD  holds  in the small-$x$ limit~\cite{Li:2019sin,Li:2019yzy,Sun:2020ygb}. The linear polarization of photons can be probed via a $\cos 4\phi$ azimuthal asymmetry in di-lepton production~\cite{Pisano:2013cya}. This asymmetry was measured in the ultra-peripheral collision (UPC) process by the STAR collaboration~\cite{Adam:2019mby} and turns out to be in a very good agreement with the theoretical expectation~\cite{Li:2019sin,Li:2019yzy}. This strongly indicates that the small-$x$ photons are fully linearly polarized. However, it has been found that the simple relation presented in Eq.\ref{linearly} is spoiled by TMD evolution~\cite{Boer:2017xpy}.    The purpose of the current work is to investigate how the relation Eq.\ref{todd} is affected by TMD evolution. To this end, we evolve T-odd gluon TMDs to high energy scales from some initial scale, where the relation  is expected to hold.  We use the expectation value of the spin dependent odderon computed in the diquark model~\cite{Szymanowski:2016mbq} as the input for the  T-odd gluon TMDs at the initial scale.

 It has been verified in a sequence of papers~\cite{Mueller:2013wwa,Zhou:2016tfe,Xiao:2017yya,Zhou:2018lfq} that the unpolarized small-$x$ gluon TMD
 satisfies the standard  Collins-Soper equation and the renormalization  group equation which hold in the
 moderate or large $x$ region. By solving the evolution equations, all large logarithms arise in higher
 order calculation can be resummed into a Sudakov factor. Such joint resummation formalism has been applied in  phenomenological studies~\cite{Zheng:2014vka,vanHameren:2014ala,Boer:2017xpy,vanHameren:2019ysa,Stasto:2018rci,Marquet:2019ltn} in various contexts. One expects that a similar analysis applies to the  polarization dependent cases. As a result, in the Collins-2011 scheme, the evolved gluon TMDs take the forms,
 \begin{eqnarray}
  f_{1}^g (x,\bm{k}_\perp^2,\mu=Q)
 &=& \int \frac{ d |\bm{b}_{\perp}|}{2\pi} ~|\bm{b}_{\perp}| J_0(|\bm{k}_{\perp}|| \bm{b}_{\perp}|)~
 e^{-S((\mu_b^2,Q^2)}~
  \tilde f_{1}^g(x,\bm{b}_{\perp*}^2),
\\
   f_{1T}^{\perp g } (x,\bm{k}_\perp^2,\mu=Q)&=& \frac{1}{|\bm{k}_{\perp}|}
  \int\frac{ d |\bm{b}_{\perp}|}{2\pi} |\bm{b}_{\perp}|
  J_1(|\bm{k}_{\perp}||\bm{b}_{\perp}|)~ e^{-S(\mu_b^2,Q^2)}~
    \tilde f_{1T}^{\perp g }(x,\bm{b}_{\perp*}^2), \label{f1t}
\\
h_{1}^{ g}(x,\bm{k}_\perp^2,\mu=Q)&=&\frac{1}{|\bm{k}_{\perp}|}
  \int \frac{ d |\bm{b}_{\perp}|}{2\pi} |\bm{b}_{\perp}|
  J_1(|\bm{k}_{\perp}||\bm{b}_{\perp}|)~e^{-S((\mu_b^2,Q^2)}~
    \tilde h_{1}^{ g}(x, \bm{b}_{\perp*}^2), \label{h1t}
\\
 h_{1T}^{\perp g}(x,\bm{k}_\perp^2,\mu=Q)&=&\frac{1}{|\bm{k}_{\perp}|^3}
  \int \frac{ d |\bm{b}_{\perp}|}{2\pi} |\bm{b}_{\perp}|
  J_3(|\bm{k}_{\perp}||\bm{b}_{\perp}|)~e^{-S((\mu_b^2,Q^2)}~
    \tilde h_{1T}^{\perp g}(x, \bm{b}_{\perp*}^2),
\end{eqnarray}
where $\mu_b=2 e^{-\gamma_E}/|\bm{b}_{\perp}| $,
the standard rapidity parameter $\zeta$ is chosen identical to the renormalization scale $\mu_b$ and 
not shown here.  The unpolarized gluon TMD and the T-odd TMDs  in $\bm{b}_\perp$
space are given by \cite{Boer:2011xd},
\begin{eqnarray}
   \tilde f_{1}^g(x,\bm{b}_{\perp}^2) &=&  2\pi \int d |\bm{l}_{\perp}| \, |\bm{l}_{\perp}| J_0(|\bm{b}_{\perp}||\bm{l}_\perp|)
  f_{1}(x,\bm{l}_\perp^2),
 \label{eqn:xf1DPk}
\\
  \tilde f^{\perp g}_{1T}(x,\bm{b}_{\perp}^2) &=&  2\pi \int d |\bm{l}_{\perp}|\, \bm{l}_\perp^2
J_1(|\bm{b}_{\perp}||\bm{l}_\perp|) ~  f^{\perp }_{1T}(x,\bm{l}_\perp^2), 
\label{eqn:xh1DPk}
\\
  \tilde h^{ g}_{1}(x,\bm{b}_{\perp}^2) &=& 2\pi \int d |\bm{l}_{\perp}| \,  \bm{l}_\perp^2
J_1(|\bm{b}_{\perp}||\bm{l}_\perp|) ~    h^{ g}_{1}(x,\bm{l}_\perp^2),
\\
\tilde h^{ \perp g}_{1T}(x,\bm{b}_{\perp}^2) &=&  2\pi \int d |\bm{l}_{\perp}| \,  \bm{l}_{\perp}^4
J_3(|\bm{b}_{\perp}||\bm{l}_\perp|) ~    h^{ \perp g}_{1T}(x,\bm{l}_\perp^2).
\end{eqnarray}

\begin{figure}[htpb]
\includegraphics[angle=0,scale=0.5]{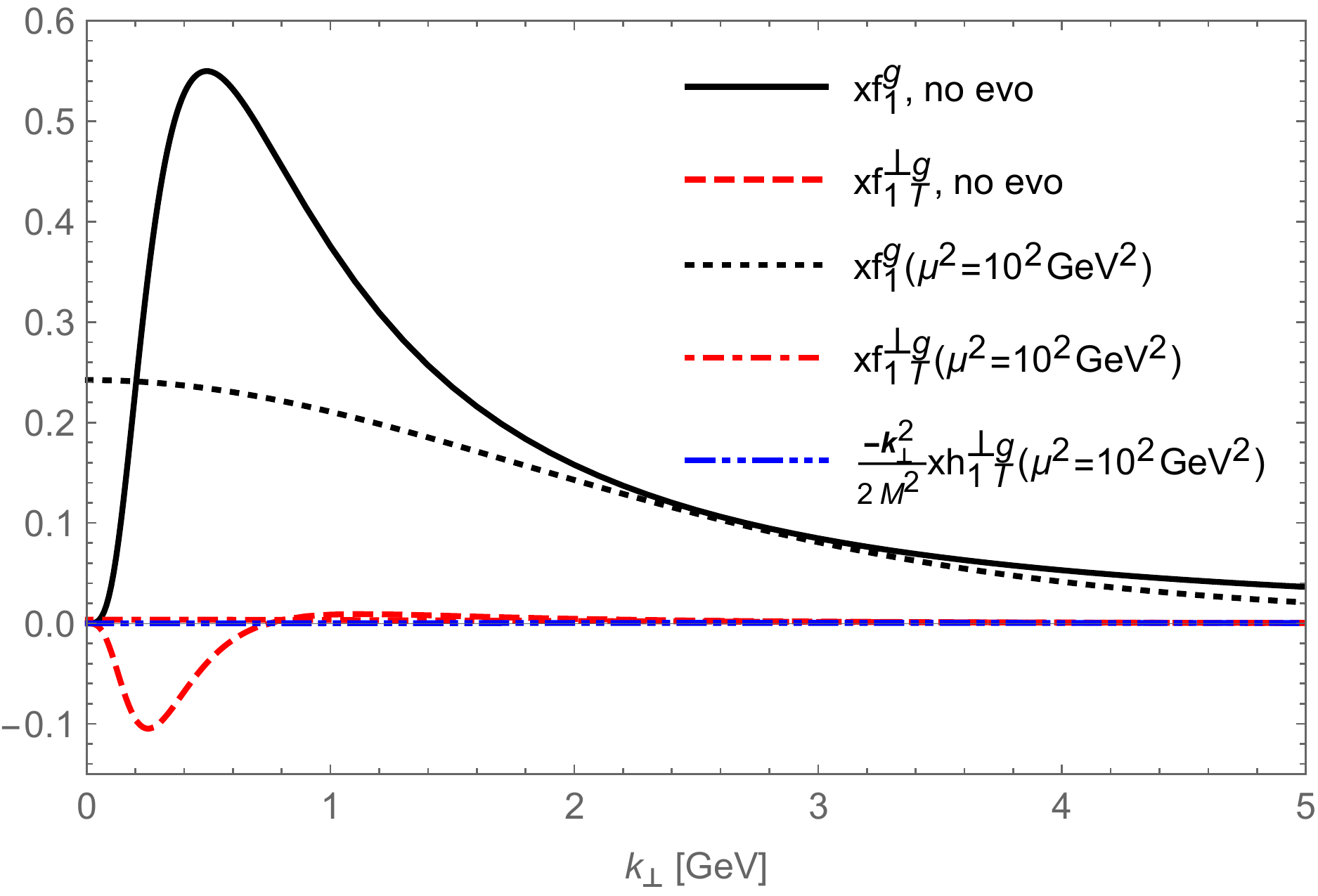}
\caption{The considered gluon TMDs at different scales at $x$=0.01, using the diquark model result as the initial condition.} \label{figs:gluonTMDs_1}
\end{figure}

\begin{figure}[htpb]
\includegraphics[angle=0,scale=0.5]{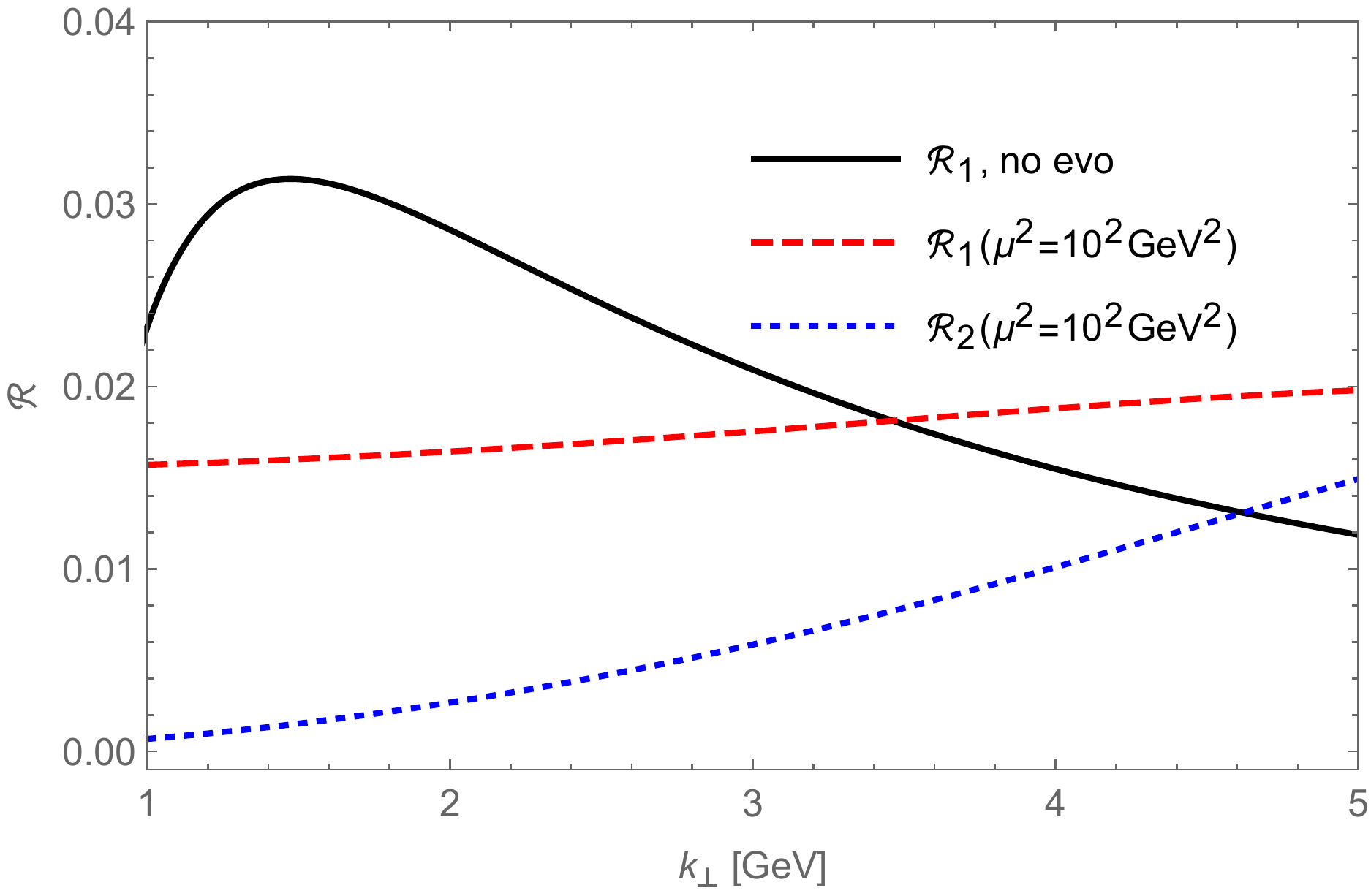}
\caption{The ratios ${\cal R}_1$ and ${\cal R}_2$ evolved to the scale $\mu$=10 GeV for $x$=0.01 in comparison to the tree level model input that satisfies ${\cal R}_1={\cal R}_2$.}.
 \label{figs:gluonTMDs_ratio}
\end{figure}

Although the energy dependence of 
  gluon TMDs is perturbatively calculable, gluon TMDs at the initial scale have to be determined either from model calculations or by fitting to experimental data.  
There are three model calculations for the spin dependent odderon, i.e.\ T-odd gluon TMDs, available in the literature~\cite{Zhou:2013gsa,Szymanowski:2016mbq,Dumitru:2021tqp}.  Our numerical results show that the spin dependent odderon computed from the MV model~\cite{Zhou:2013gsa} is very small and would not lead to any measurable effects.  As the MV model is well justified only for a large nucleus target, the MV model calculation for a proton target should probably not be taken too seriously.  In the following numerical estimations,  we use the diquark model results as input for both the unpolarized gluon TMDs and the spin dependent gluon TMDs at the initial scale~\cite{Szymanowski:2016mbq}.  In this model the quark-diquark system in the proton forms the source of the small-$x$ gluons, leading to the following unpolarized gluon distribution:
\begin{eqnarray}
\!\!\!\!\!\! xf_1^g(x,\bm{k}_\perp^2) =\frac{2 \lambda_s^2 C_F N_c \alpha_s }{(2\pi)^5 \bm{k}_\perp^2 } \!\! \int \! dz d^2\bm{x}_\sperp \,  \bar z \! \left[  \left( Mz+m_q \right)^2K_0^2(\tilde M |\bm{x}_\sperp|) +\tilde M^2 K_1^2(\tilde M |\bm{x}_\sperp|) \right ] (1\!-\!e^{-i\bm{x}_\sperp \cdot \bm{k}_\perp}) (1\! -\! e^{i\bm{x}_\sperp \cdot \bm{k}_\perp}).
\end{eqnarray}
Here $K_1$ and $K_0$ are the modified Bessel functions of the second kind, $\bar{z}=1-z$, $M$ stands for proton mass, and ${\tilde M}^2=\bar z m_q^2 +zm_s^2-z\bar z M^2$ with the valence quark mass and the scalar diquark mass  taken to be $m_q=0.3$ GeV and $m_s=0.8$ GeV, respectively. The strong coupling constant is fixed to be $\alpha_s=0.3$ in the diquark model calculation. We further  fix the proton-quark-scalar diquark effective coupling constant  $\lambda_s$ 
by requiring: $\pi \int_0^{\mu^2} d \bm{k}_\perp^2 x f_1^g(x, \bm{k}_\perp^2) = x G(x, \mu )$ with $x G(x, \mu )$ being the standard gluon PDF taken from MSTW 2008 LO PDF set, and at $x \sim 0.01$ that is the typical kinematical region we consider in the next section. With this we obtain $\lambda_s^2 \simeq 153$ at $\mu=1.6$ GeV which is the lowest scale in $S_{NP}$ which will be introduced  shortly. This corresponds to the ``no evo" curves in Fig.\ref{figs:gluonTMDs_1} and \ref{figs:gluonTMDs_ratio}. For the evolved curves $\lambda_s^2$ is dependent on $\mu$. An infrared cutoff is imposed by replacing $\frac{1}{\bm{k}_\perp^2}$ with  $\frac{1}{\bm{k}_\perp^2+\Lambda_{QCD}^2}$ when Fourier transforming the above expression to  impact parameter space.  The spin dependent gluon TMDs in the diquark model read~\cite{Szymanowski:2016mbq},
\begin{eqnarray}
&&\!\!\! xf_{1T}^{\perp g}(x,\bm{k}_\perp^2)=xh_{1}^{ g}(x,\bm{k}_\perp^2)
=\frac{-\bm{k}_{\perp}^2}{2 M^2}x h_{1T}^{\perp g}(x,\bm{k}_\perp^2)
=\frac{  i  \lambda_s^2M\alpha_s^2 C_F (N_c^2-4)}{2 (2 \pi)^5  \bm{k}_\perp^2} \int\!dz \,d^2\bm{x}_\sperp\, \bar z\,\left( Mz+m_q  \right) \tilde M \, \frac{\bm{k}_{\perp} \cdot \bm{x}_\sperp}{\bm{k}_\perp^2 |\bm{x}_\sperp|}  \\ && \times K_0(\tilde M |\bm{x}_\sperp|)
K_1(\tilde M|\bm{x}_\sperp|) (1\!-\! e^{\! -i\bm{k}_{\perp} \! \cdot \bm{x}_\sperp})\! \int_0^1 \! \frac{da}{a\bar a} \! \left \{\! 
1+e^{i\bm{k}_{\perp} \cdot \bm{x}_\sperp}\! -\! \sqrt{a\bar a \bm{x}_\sperp^2 \bm{k}_\perp^2} K_1(\sqrt{a\bar a \bm{x}_\sperp^2\bm{k}_\perp^2} \
) \left( \!e^{ia\bm{k}_{\perp} \!  \cdot \bm{x}_\sperp} +e^{i\bar a \bm{k}_{\perp} \! \cdot \bm{x}_\sperp}\! \right ) \! \right \}\! +c.c.\; \nonumber 
\end{eqnarray}
It is interesting to notice that the spin dependent odderon can not be related to the derivative of the unpolarized gluon TMD in the di-quark model  as it  can be  in the MV model~\cite{Zhou:2013gsa}.

The standard treatment for the non-perturbative part applies to the Sudakov factor $S(\mu_b^2,Q^2)$ which at one-loop order reads,
\begin{eqnarray}
S(\mu_b^2,Q^2)=  S_A(\mu_{b*}^2,Q^2) +S^{NP}(b_\perp^2, Q^2),
\end{eqnarray}
with 
\begin{eqnarray}
 S_A(\mu_{b*}^2,Q^2) =
\frac{C_A}{2\pi} \int^{Q^2}_{\mu_{b*}^2} \frac{d\mu^2}{\mu^2} \alpha_s(\mu) \left[ \ln
\frac{Q^2}{\mu^2} - \frac{11-2n_f/C_A}{6} \right ],
\end{eqnarray}
where $\mu_{b*}^2$ is defined as $\mu_{b*}^2=4e^{-2\gamma_E}/b_{\perp*}^2$, with $b_{\perp*}$ given by
$b_{\perp *}=\frac{b_{\perp}}{\sqrt{1+b_{\perp}^2/b^2_{\max}}} $ and $b_\perp = |\bm{b}_\perp|$. A regulator which allows us to smoothly match  large transverse momentum behavior is introduced~\cite{Boer:2014tka,Boer:2015uqa,Collins:2016hqq}
\begin{equation}
\mu_b^\prime(Q^2)= \frac{1}{\sqrt{1/\mu_{b*}^2  +1/Q^2 }}.
\end{equation}
The parametrization for the non-perturbative Sudakov factor $S^{NP}(b_\perp^2,Q^2)$ is taken from \cite{Aybat:2011zv} and Casimir scaled to apply to gluons rather than quarks: 
\begin{equation}
S_{g}^{NP}(b_{\perp}^2,Q^2)=\frac{C_A}{C_F}
\frac{1}{2}\left( g_1+g_2 \ln \frac{Q}{2 Q_0} + 2 g_1 g_3 \ln \frac{10 x x_0}{x_0+x} \right)
b_{\perp}^2,
\end{equation}
with $b_{\max}=1.5\, \text{GeV}^{-1},~ g_1=0.201\, \text{GeV}^2, ~ g_2=0.184\, \text{GeV}^2,~g_3=-0.129,~x_0=0.009,~Q_0=1.6\, \text{GeV}$. In our numerical estimation,  we use the one-loop running coupling constant $\alpha_s$, with $n_f=3$ and $\Lambda_{\text{QCD}} = 216~ \text{MeV}$. Here we assume for simplicity that the unpolarized gluon TMD and the T-odd gluon TMDs share the same non-perturbative part of the Sudakov factor, which is not necessarily the case though.

With the expressions introduced above, we are now ready to evolve the T-odd gluon TMDs computed in the diquark model  to a high scale. Judging from Eq.\ref{f1t}
and Eq.\ref{h1t}, the gluon Sivers function and the gluon TMD $h_{1}^{g}$ evolve in exactly the same way and thus remain identical at arbitrary energy scale, so $h_1^g$ will not be shown in the figures.
In Fig.\ref{figs:gluonTMDs_1} we present  the unpolarized gluon distribution $x f_1^g$,
 the gluon Sivers function $x f_{1T}^{\perp g}$,  and $x h_{1T}^{\perp g}$ as  function
 of $k_\perp$ (with $k_\perp = |\bm{k}_\perp|$) at the initial scale $\mu=1.6$ GeV and the scale  $\mu=\text{10 GeV}$.
 One can see that $x f_{1T}^{\perp g}$ and $x h_{1T}^{\perp g}$ are much smaller than $x f_1^g$. Since the azimuthal asymmetries are proportional to the ratios ${\cal R}_1(\mu^2)=\frac{x f_{1T}^{\perp g}(\mu^2)}{x f_{1}^{g}(\mu^2)}$, ${\cal R}_2(\mu^2) = \frac{-\frac{\bold{k}_\perp^2}{2 M^2} x h_{1T}^{\perp g}(\mu^2)}{x f_{1}^{g}(\mu^2)}$, we  directly plot these   ratios at the initial scale and the scale $\mu=\text{10 GeV}$ as  function of  $k_\perp$ in Fig.\ref{figs:gluonTMDs_ratio}. The $k_\perp$ dependent behavior of both ratios ${\cal R}_1$ and ${\cal R}_2$ changes quite substantially after performing energy evolution. ${\cal R}_1$ is about $2\%$ at $\mu=\text{10 GeV}$ and is only very mildly  dependent of $k_\perp$. The  ${\cal R}_2$ rises with increasing  $k_\perp$, but remains quite small at low transverse momentum.
We note that although there is a large model uncertainty and we do not expect the diquark model to be very realistic, the effect of the evolution {\it is} expected to reflect the actual situation, namely that ${\cal R}_1$ starts to differ from  ${\cal R}_2$ under TMD evolution quite quickly.

\section{Asymmetries in virtual photon-jet production in polarized $pp$ collisions}
Azimuthal asymmetries in virtual photon-jet production in polarized $pp$ collisions allow to
probe the T-odd gluon TMDs discussed in the previous section. The dominant partonic process is
\begin{equation}
 q(x_q \bar P) + g(x_gP+k_{\perp})  \to \gamma^*(p_1)+q(p_2)
\end{equation}
where the quark comes from the unpolarized projectile,
 and the incoming gluon  from the transversely polarized target.
 The virtual photon and quark in the final state are produced in the forward region of the unpolarized
 projectile. In this kinematical region, transverse momentum carried by the incoming gluon is much larger than
 that of the incoming quark. As such, a hybrid approach in which the polarized target is treated as a CGC, while
 on the side of dilute projectile one uses the ordinary integrated parton distribution functions is justified.
 The hybrid approach was widely used~\cite{Gelis:2002ki} to compute both spin averaged observables  and polarization dependent observables in the forward/backward region of $pA$/$pp$ collisions.

In order for TMD factorization to be valid, we restrict ourselves to the correlation limit where each of
the produced particles transverse momentum $\bm{p}_{1\perp}$ or $\bm{p}_{2\perp}$ is much larger
than their sum $\bm{k}_\perp=\bm{p}_{1\perp}+\bm{p}_{2\perp}$, i.e.\ the probed gluon transverse momentum.
In the correlation limit, one has,
 \begin{eqnarray}
&&\bm{P}_\perp \equiv \frac{\bm{p}_{1\perp}-\bm{p}_{2\perp}}{2} \sim  \bm{p}_{1\perp} \sim -\bm{p}_{2\perp}
\\
&&k_\perp=|\bm{p}_{1\perp}+\bm{p}_{2\perp}| \ll P_\perp
 \end{eqnarray}
 with $P_\perp = |\bm{P}_\perp|$. Here, $P_\perp^2$ serves as an additional hard scale required by TMD factorization. In the reference frame in
 which azimuthal angles are measured with respect to the gluon transverse momentum vector, i.e.\ $\phi_k=0$,
 the azimuthal angles of  the vectors $\bm{S}_\perp$, $\bm{P}_\perp$ are denoted by  $\phi_S$, $\phi_P$, respectively.
 The calculation of the polarization dependent differential cross section
 proceeds along the same lines of Ref.~\cite{Boer:2017xpy}. We first apply the Eikonal approximation and sum all order gluon
re-scatterings into the Wilson lines. Combining the amplitude with its conjugate part, two Wilson lines form
a closed Wilson loop. Due to the different charge parity properties, the real part of the Wilson loop contributes
to the unpolarized cross section, while the imaginary part of the dipole amplitude is responsible for the spin
dependent contributions. The next step is to isolate the leading power
part of the hard coefficients by  Taylor expanding them in terms of powers of $k_\perp/P_\perp$ and
convert $k_\perp^i$ into $\partial_\perp^i$ that acts on the Wilson lines by partial integration.
After having done so, the derivative of the Wilson loop can be related to the gluon TMD matrix element.
One eventually recovers the differential cross section in TMD factorization, which reads,
\begin{eqnarray} 
\frac{d\sigma^{p^\uparrow p\to \gamma^*qX}}{dP.S}&=&\sum_q  x_q  f_1^q(x_q)
 \left \{ H_{UU} \left [x f^g_{1}(x,\bm{k}_{\perp}^2)+ \sin (\phi_S)
\frac{|\bm{k}_\perp|}{M} x f_{1T}^{\perp g}(x,\bm{k}_{\perp}^2)\right ]  \right.
 \nn \\ 
 &+& \left.  H_{UT}    \cos(2\phi_P)  \frac{|\bm{k}_\perp|^2}{2M^2} x h_1^{\perp g} (x,\bm{k}_{\perp}^2) 
  +    \frac{1}{2} H_{UT}  \sin(2\phi_P-\phi_S)  \frac{|\bm{k}_\perp|}{M} x h_{1}^g (x,\bm{k}_{\perp}^2)\right.\nn \\  
  &+&  \left.  \frac{1}{2} H_{UT} \sin(2\phi_P +\phi_S)
   \frac{|\bm{k}_\perp|}{M} \frac{|\bm{k}_\perp|^2}{2 M^2} x h_{1T}^{\perp g}(x,\bm{k}_{\perp}^2)  \right \} ,
\end{eqnarray}
where $f_1^q(x_q)$ is the quark collinear PDF of the proton and the hard coefficients are given by,
\begin{eqnarray}
H_{UU}&=&\frac{\alpha_s \alpha_{em} e_q^2}{N_c \hat s^2}
\left [ -\frac{\hat s}{\hat t}-\frac{\hat t}{\hat s}- \frac{2Q^2\hat u}{\hat s \hat t} \right ],
\\
H_{UT}&=&\frac{\alpha_s \alpha_{em} e_q^2}{N_c \hat s^2}\frac{-2Q^2 \hat u }{\hat s \hat t} .
\end{eqnarray}

In the above expressions, the phase space factor is defined as $d P.S=dy_J dy_{\gamma^*} d^2\bm{P}_\perp
d^2 \bm{k}_\perp$, where $y_J$ and $ y_{\gamma^*}$ are the rapidities of the produced jet and the
virtual photon, respectively. $Q^2$ and $z$ are the virtual photon invariant mass and  the
longitudinal momentum fraction of the incoming quark carried by the virtual photon, respectively.

Similar to the case of asymmetries in di-jet production in the SIDIS process~\cite{Boer:2016fqd}, there are three independent azimuthal modulations, each of which is related to a different T-odd gluon TMD. Moreover, one observes that
in close analogy to the $\cos 2\phi$ asymmetry induced by the linearly polarized gluon distribution  in unpolarized $pA$/$pp$ collisions, the size of the azimuthal asymmetries given by the last two terms is proportional  to $Q^2$ and thus vanish for real photon production. In order to single out separate angular dependence,  we define the following azimuthal moments,  
 \begin{equation}
\left\langle \sin(\phi_S)\right\rangle
 \equiv
\frac{\int d \phi_P \; d\phi_S \;\sin
( \phi_S) \; \left[d \sigma(\phi_P, \phi_S) -
  d \sigma(\phi_P,\phi_S +\pi)\right] }{\int d \phi_P \; d\phi_S\;
\left[d \sigma(\phi_P, \phi_S) + d \sigma(\phi_P,\phi_S +\pi)\right] }
\end{equation}
and similarly,
\begin{equation}
\left\langle\sin(2 \phi_P+\phi_S) \right\rangle
 \equiv
\frac{\int d \phi \; d\phi_S \;\sin
(2 \phi_P+ \phi_S) \; \left[d \sigma(\phi_P, \phi_S) -
  d \sigma(\phi_P,\phi_S +\pi)\right] }{\int d \phi_P\; d\phi_S\;
\left[d \sigma(\phi_P, \phi_S) + d \sigma(\phi_P,\phi_S +\pi)\right] }
\end{equation}

\begin{equation}
\left\langle \sin(2 \phi_P-\phi_S) \right\rangle
 \equiv
\frac{\int d \phi \; d\phi_S \;\sin
(2 \phi_P-\phi_S) \; \left[d \sigma(\phi_P, \phi_S) -
  d \sigma(\phi_P,\phi_S +\pi)\right] }{\int d \phi_P\; d\phi_S\;
\left[d \sigma(\phi_P, \phi_S) + d \sigma(\phi_P,\phi_S +\pi)\right] }
\end{equation}
Using the derived spin dependent cross section, it is easy to express the azimuthal moments in terms of the unpolarized gluon TMD and the corresponding T-odd gluon TMDs,
\begin{eqnarray}
\left\langle \sin(\phi_S)\right\rangle
=\frac{\sum_q  \int dP.S ~d |\bm{b}_\perp| |\bm{b}_\perp| x_q  f_1^q(x_q, \mu_b^2)  e^{-S}\, \frac{1}{|\bm{k}_\perp|}  J_1(|\bm{k}_\perp| |\bm{b}_\perp|) x \tilde{f}_{1T}^{\perp g } (x,\bm{b}_{\perp}^2) \frac{|\bm{k}_\perp|}{M} H_{UU}}{\sum_q  \int dP.S ~d |\bm{b}_\perp| |\bm{b}_\perp| x_q  f_1^q(x_q, \mu_b^2)  e^{-S}\,  J_0(|\bm{k}_\perp| |\bm{b}_\perp|) 2 x f_1^g(x,\bm{b}_{\perp}^2) H_{UU}}
\end{eqnarray}
\begin{equation}
\left\langle\sin(2 \phi_P+\phi_S) \right\rangle
=\frac{\sum_q  \int dP.S ~d |\bm{b}_\perp| |\bm{b}_\perp| x_q  f_1^q(x_q, \mu_b^2)  e^{-S} \, \frac{1}{|\bm{k}_\perp|^3}  J_3(|\bm{k}_\perp| |\bm{b}_\perp|) x \tilde{h}_{1T}^{\perp g}(x,\bm{b}_{\perp}^2) \frac{|\bm{k}_\perp|}{M}\frac{|\bm{k}_\perp|^2}{2M^2}  \frac{1}{2}H_{UT}}{\sum_q  \int dP.S ~d |\bm{b}_\perp| |\bm{b}_\perp| x_q  f_1^q(x_q, \mu_b^2)  e^{-S}\,  J_0(|\bm{k}_\perp| |\bm{b}_\perp|) 2 x f_1^g(x,\bm{b}_{\perp}^2) H_{UU}} 
\end{equation}
\begin{eqnarray}
\left\langle \sin(2 \phi_P-\phi_S) \right\rangle
=\frac{\sum_q  \int dP.S ~d |\bm{b}_\perp| |\bm{b}_\perp|  x_q  f_1^q(x_q, \mu_b^2)  e^{-S}  \,  \frac{1}{|\bm{k}_\perp|}  J_1(|\bm{k}_\perp| |\bm{b}_\perp|) x \tilde{h}_{1}^{g } (x,\bm{b}_{\perp}^2) \frac{|\bm{k}_\perp|}{M} \frac{1}{2}H_{UT}}{\sum_q  \int dP.S ~d |\bm{b}_\perp| |\bm{b}_\perp| x_q  f_1^q(x_q, \mu_b^2)  e^{-S}\,  J_0(|\bm{k}_\perp| |\bm{b}_\perp|) 2 x f_1^g(x,\bm{b}_{\perp}^2) H_{UU}}
\end{eqnarray}
The  Sudakov factor $S$ is given by~\cite{Hatta:2021jcd}
\begin{eqnarray}
S=\!\int_{\mu_b^2}^{P_\perp^2} \! \frac{d\mu^2}{\mu^2} \frac{\alpha_s(\mu)}{2\pi} \left \{ (C_A+C_F)\ln \frac{\hat s}{\mu^2}- C_A \left (\frac{11}{6}\!-\!\frac{n_f}{9} \right ) -\frac{3C_F}{2}+(C_A\!-\!C_F) (y_J\!-\!y_{\gamma^*})+C_F \ln \frac{1}{R^2} \right \} +S_{NP}
\end{eqnarray}
 where $S_{NP}$ is the nonperturbative part of the Sudakov factor. The term $C_F \ln \frac{1}{R^2}$  in the above formula arises from the final state soft gluon emissions along the jet direction~\cite{Banfi:2003jj,Sun:2014gfa,Catani:2014qha,Catani:2017tuc,Liu:2018trl,Chien:2019gyf,Chien:2020hzh,Hatta:2020bgy}.
 $R$ is the  produced jet radius which is chosen to be 0.4 in the following numerical estimations. Notice that the final state gluon radiation can lead to $\cos \phi$, $\cos 2 \phi$, $\cos 3\phi$, $\ldots$ azimuthal modulations in this process~\cite{Hatta:2021jcd}, where $\phi$ is the angle between the two transverse momenta $\bm{P}_\perp$ and $ \bm{k}_\perp$.

\begin{figure}[h]
\includegraphics[angle=0,scale=0.4]{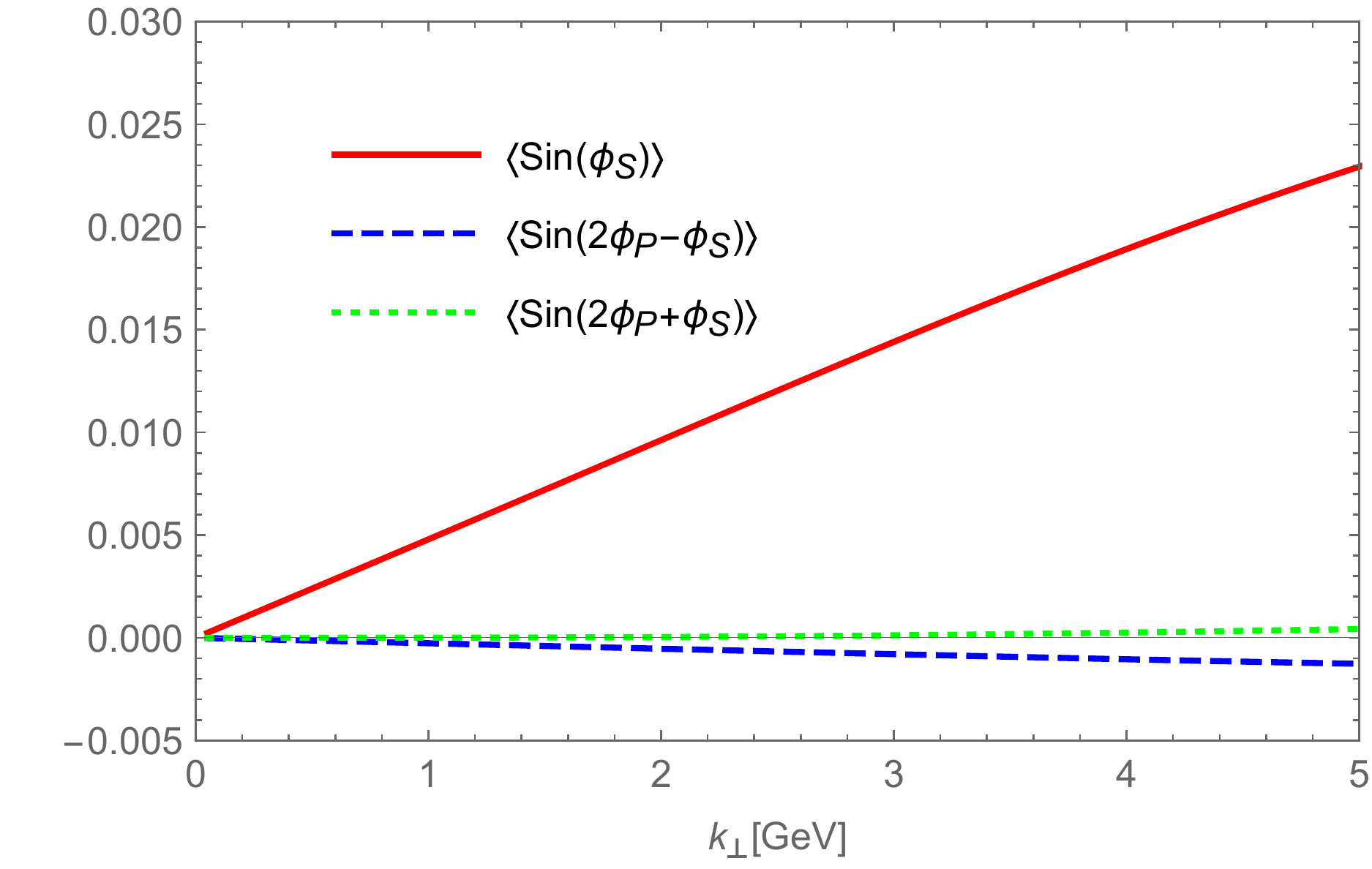}
\includegraphics[angle=0,scale=0.4]{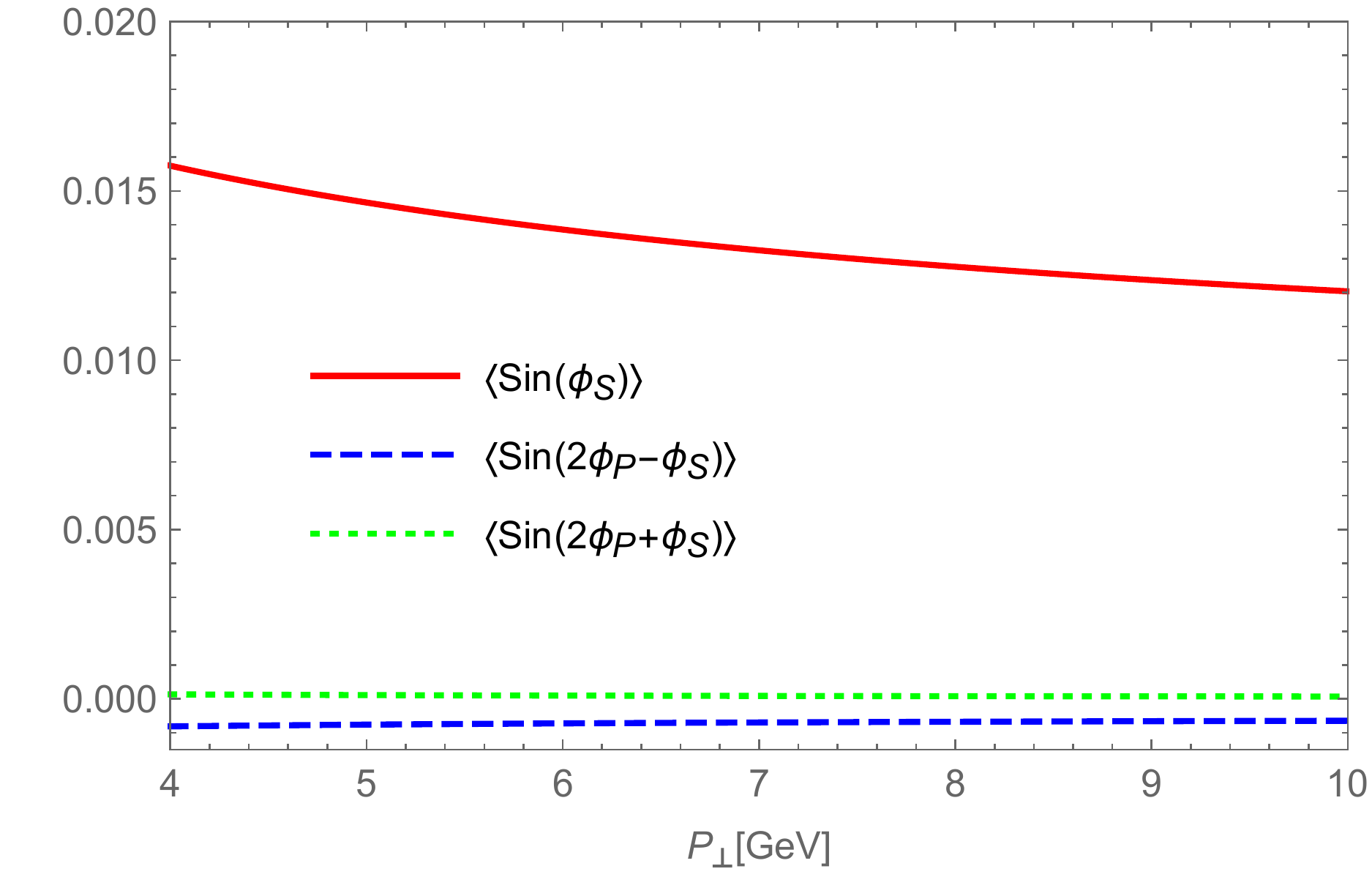}
\caption{Estimates of the azimuthal asymmetries as function of $k_\perp$ for $P_\perp=10$ GeV (left panel), and function of $P_\perp$ for $k_\perp=2.5$ GeV (right panel) at $\sqrt s=500$ GeV. The jet and virtual photon rapidities 
$y_J$, $y_\gamma^*$ are integrated over the regions $[-1.5,-0.5], [-3,-2]$, respectively. The invariant mass of the virtual photon $Q^2$ is integrated over the region [30, 80] $\text{GeV}^2$ in the left figure, and set to be $\frac{1}{2} P_{\perp}^2$ in the right one.} 
\label{figs:azimuthal_asymmetry}
\end{figure}

 The numerical model results for the asymmetries  in different kinematic regions are presented in Fig.\ref{figs:azimuthal_asymmetry}. The largest asymmetry is the $\left\langle \sin(\phi_S)\right\rangle$ which reaches two percent. It grows with increasing $k_\perp$ in the TMD region at low transverse momentum $k_\perp < P_\perp/2$. The $k_\perp$ dependence of $\left\langle \sin(2 \phi_P-\phi_S) \right\rangle$ 
 is similar to that of $\left\langle \sin(\phi_S)\right\rangle$, while the magnitude of  $\left\langle \sin(2 \phi_P-\phi_S) \right\rangle$ is suppressed due to the smaller associated hard part. In fact, the difference between the curves for $\left\langle \sin(\phi_S)\right\rangle$ and $\left\langle \sin(2 \phi_P-\phi_S) \right\rangle$ solely reflects the difference between $H_{UT}$ and $H_{UU}$, while the difference between the curves for $\left\langle \sin(2 \phi_P-\phi_S) \right\rangle$ and $\left\langle\sin(2 \phi_P+\phi_S) \right\rangle$ solely reflects the difference due to the TMD evolution. Unfortunately the predicted curves for $\left\langle \sin(2 \phi_P-\phi_S) \right\rangle$ and $\left\langle\sin(2 \phi_P+\phi_S) \right\rangle$ are found to be very small in this model calculation and it would pose a big challenge to measure them experimentally. However, since our model calculation may be far from realistic, it would nevertheless be worth measuring all three angular dependences at RHIC, because at present it is the only known way to investigate the differences between the three T-odd dipole gluon TMDs experimentally. Although the dipole gluon Sivers TMD can also be measured through the single spin left-right asymmetry $A_N$ in charged hadron production in the backward region of $p^\uparrow p$ collisions~\cite{Boer:2015pni}, that process would not allow to access the other two T-odd dipole gluon TMDs and therefore, also not their different TMD evolutions.

Our numerical results indicate that the Sivers type asymmetry will be the largest of the three asymmetries. Once the dipole gluon Sivers TMD is measured through this asymmetry or through the left-right asymmetry $A_N$ in backward charged hadron production, it will be possible to estimate the size of the other two smaller asymmetries in virtual photon-jet production with more certainty.

\section{Conclusions}
 In summary, we study the energy (TMD) evolution of three T-odd gluon TMDs relevant for the single spin asymmetries.  All  three dipole gluon TMDs are related to the spin dependent odderon in the small-$x$ limit, and are shown to be identical at the tree level. Although this relation persists under small-$x$ evolution, it is violated by TMD  evolution due to the different polarization tensor structures involved. Recent theoretical developments have confirmed that both double and single Sudakov logs arising in the small-$x$ limit can be summed to all orders in the conventional TMD resummation formalism. We thus carry out a detailed numerical studies of the evolved T-odd gluon TMDs in the Collins-2011 scheme using the diquark model expressions as the initial conditions.  We found that the ratios between the unpolarized gluon TMD and the different polarized gluon TMDs exhibit strong energy  dependences.  It would be interesting to test these predictions by experimentally studying T-odd gluon TMDs in virtual photon-jet production in polarized $pp$ collisions, where different azimuthal modulations can serve as analyzers of the different polarized gluon TMDs. 
 
 Our numerical model results suggest that the azimuthal asymmetry induced by the gluon Sivers function is the largest one, whereas the asymmetries generated by other two T-odd gluon TMDs are much smaller. The absolute magnitude of the asymmetries critically depends on the model input at the initial scale, which is very uncertain. However, the relative sizes of the different asymmetries are more or less independent of the model setup, as they are mainly determined by the evolution effect and the different hard factors. Once the Sivers type asymmetry has been measured, it will be possible to make more realistic predictions for the other two azimuthal asymmetries in the virtual photon-jet process. We stress that at present this process offers the only known way to investigate the differences between the three T-odd dipole gluon TMDs experimentally. This makes it especially interesting to be studied at RHIC. Observation of any of these single spin asymmetries would constitute a clear first signal of the spin dependent odderon.

\begin{acknowledgments}
Jian Zhou has been supported by the National Natural Science Foundations of China under Grant No.\ 12175118. Ya-jin Zhou has been supported by  the  Natural Science Foundation of Shandong Province under Grant No.\ ZR2020MA098. 
\end{acknowledgments}

\bibliography{ref.bib}

\end{document}